\documentclass[envcountsame]{llncs}
\usepackage{tikz}
\usetikzlibrary{snakes}

\newcommand{\systemname}[1]{\emph{#1}}
\newcommand{\alphabet}{\mathcal{A}}
\newcommand{\atom}[1]{atom(#1)}
\newcommand{\head}[1]{head(#1)}
\newcommand{\body}[1]{body(#1)}
\newcommand{\dneg}[0]{\sim\!\!}
\newcommand{\nodelabel}[2]{#1_{\scriptscriptstyle\!#2}}
\newcommand{\dneglabel}[2]{{\scriptstyle\sim\!}\nodelabel{#1}{#2}}

\begin{document}

\title{Symmetry-breaking Answer Set Solving}

\author{Christian Drescher\inst{1} \and Oana Tifrea\inst{2} \and Toby Walsh\inst{3}}

\institute{Vienna University of Technology, Austria
\and Free University of Bozen-Bolzano, Italy
\and NICTA and University of New South Wales, Australia}

\maketitle

\begin{abstract}
In the context of Answer Set Programming, this paper investigates symmetry-breaking to eliminate symmetric parts of the search space and, thereby, simplify the solution process.
We propose a reduction of disjunctive logic programs to a coloured digraph such that permutational symmetries can be constructed from graph automorphisms.
Symmetries are then broken by introducing symmetry-breaking constraints.
For this purpose, we formulate a preprocessor that integrates a graph automorphism system. Experiments demonstrate its computational impact.
\end{abstract}

\section{Introduction}
Answer Set Programming (ASP; \cite{baral03}) has been shown to be a useful approach for knowledge representation and non-monotonic reasoning in various applications that include difficult combinatorial search, among them graph theoretic problems, planning, model checking, and problems from bioinformatics. ASP combines an expressive but simple modelling language, able to encode all search problems within the first three levels of the polynomial hierarchy, with high-performance solving capacities \cite{drgegrkakoossc08a}. In fact, ASP solvers have experienced dramatic improvements in their performance \cite{gekasc09b} and compete\footnote{\texttt{http://www.satcompetition.org/}} with the best Boolean Satisfiability (SAT; \cite{bihemawa09a}) solvers.

However, many combinatorial search problems exhibit symmetries which can frustrate a search algorithm to fruitlessly explore independent symmetric subspaces. Various instance families, such as the \emph{Pigeon Hole} problem, are known to require exponential time for resolution and backtracking algorithms \cite{ur87a}. Indeed, state-of-the-art ASP solvers take a very long time to solve those instances (Section \ref{sec:exp}). Once their symmetries are identified, it is possible to avoid redundant computational effort by pruning parts of the search space through symmetry-breaking.

Symmetry-breaking also addresses post-processing: Where symmetries induce equivalence classes in the solution space, symmetric solutions can be discarded. Problems like the \emph{All-interval Series} taken from the CSPLib~\cite{gewa99a} have plenty symmetric solution. However, all solutions to the original problem can be reconstructed from the answer sets under symmetry-breaking.

In this paper we break the problem of symmetry-breaking down into two parts: (1) identifying symmetries and (2) breaking the identified
symmetries. We adopt existing theoretical foundations on introducing sym\-me\-try-breaking constraints (SBC) for SAT instances in conjunctive normal form (CNF) that exhibit symmetries \cite{crgiluro96a,almasa03a,alramasa03a,sa09a}. As to SAT, the basic idea is to detect irredundant generators of the group of permutational symmetries using a reduction to coloured graph automorphism. For each such generator, an SBC is constructed and added to the original CNF formula.

The key contribution of our work is a reduction of symmetry detection for disjunctive logic programs to the automorphisms of a coloured digraph, and an ASP representation of SBC which is linear in the number of problem variables. Furthermore, we formulate Symmetry-breaking Answer Set Solving as preprocessing and demonstrate its computational impact on difficult combinatorial search problems.

The remainder of this paper is organized as follows. We start by giving the necessary background notions of ASP and group theory. In turn, Sections~\ref{sec:sd} and~\ref{sec:sb} cover symmetry detection and symmetry-breaking for ASP, respectively. In Section~\ref{sec:psb} we motivate partial symmetry-breaking, in Section~\ref{sec:exp} we empirically evaluate our approach. Section~\ref{sec:con} draws conclusions.

\section{Logical Background}
A \emph{(disjunctive) logic program} over a set of primitive propositions $\alphabet$ is a finite set of \emph{rules} $r$ of the form
\begin{equation} \label{form:rule}
a_1 ; \dots ; a_l \leftarrow b_1 , \dots , b_m, \dneg c_1 , \dots , \dneg c_n
\end{equation}
where and $a_i, b_j, c_k \in \alphabet$ are \emph{atoms} for $1 \leq i \leq l$, $1 \leq j \leq m$, and $1 \leq k \leq n$.
A \emph{literal} is an atom $a$ or its default negation $\dneg a$.
Let $\head{r} = \{a_1 , \dots , a_l\}$ be the \emph{head} of $r$ and $\body{r} = \{b_1 , \dots , b_m, \dneg c_1 , \dots , \dneg c_n\}$ the \emph{body} of $r$. For a set of literals $S$, define $S^{+} = \{a \mid a \in S\}$ and $S^{-} = \{a \mid \dneg a \in S\}$. The set of atoms occurring in a logic program $P$ is denoted by $\atom{P}$, and the set of bodies in $P$ is $\body{P} = \{ \body{r} \mid r \in P \}$. For regrouping bodies sharing the same head atom $a$, define $\body{a} = \{ \body{r} \mid r \in P,\  a \in \head{r} \}$.

The semantics of a logic program is given by its answer sets. A set $M \subseteq \alphabet$ is an \emph{answer set} of a logic program $P$ over $\alphabet$, if $M$ is a $\subseteq$-minimal model of the \emph{reduct} \cite{gellif91a}
\[
P^M = \{ \head{r} \leftarrow \body{r}^{+} \mid r \in P,\ \body{r}^{-} \cap M = \emptyset\}.
\]
A rule of form (\ref{form:rule}) can be seen as a constraint on the answer sets of a program, stating that if $b_{l+1} , \dots , b_m$ are in the answer set and none of $c_{m+1} , \dots , c_n$ are included, then one of $a_1, \dots, a_l$ must be in the set. Clearly, an answer set induces a truth assignment on the atoms in $P$.

The semantics of important extensions to logic programs, such as integrity constraints, is given through program transformations that introduce additional propositions \cite{siniso02a}.
An \emph{integrity constraint} of the form
\begin{equation} \label{form:integrity}
\leftarrow b_1 , \dots , b_m, \dneg c_1 , \dots , \dneg c_n
\end{equation}
is a short hand for a rule with an unsatisfiable head, and thus forbids its body to be satisfied in any answer set.

\begin{example} Consider the logic programs $P_1$ and $P_2$, both have two answer sets $\{a\}$ and $\{b\}$, where
\[
P_1 = \left\{
\begin{array}{r@{\ \leftarrow\ }l}
a & \dneg b\\
b & \dneg a\\
\end{array}\right\},
\qquad
P_2 = \left\{
\begin{array}{r@{\ \leftarrow\ }l}
a ; b & \\
& a, b\\
\end{array}\right\}.
\]
Observe, that $P_1$ and $P_2$ remain invariant under a swap of atoms $a$ and $b$ which is what we call a symmetry. In this work we will only deal with symmetries that can be thought of as permutations of atoms.
\end{example}

\begin{example} \label{ex:allint}
The \emph{All-interval Series problem} is to find a permutation of the $n$ integers from $0$ to $n-1$ such that the difference of adjacent numbers are also all-different. We encode the All-interval Series problem introducing propositional variables $v_{i,j}$ and $d_{k,l}$ for the $i \in 1..n$ integer variables taking values $j\in 0..(n-1)$, and $k \in 1..(n-1)$ auxiliary variables taking values $l \in 1..(n-1)$ to represent the differences between adjacent numbers, respectively. Furthermore, we require both sets of variables to have pairwise different values (all-different constraint).
\[
\begin{array}{r@{\ \leftarrow\ }l@{\qquad}l}
v_{i,0} ; v_{i,1} ; \dots ; v_{i,n-1} &  & i \in 1..n\\
& v_{i,j}, v_{i,k} & j \neq k \\
d_{i,|j-k|} & v_{i,j}, v_{i+1,k} & i \in 1..(n-1) \land j,k \in 0..(n-1)\\
& d_{i,j}, d_{i,k} & j \neq k \\
\end{array}
\]
Note that above encoding for the all-different constraint corresponds to its \emph{support encoding}~\cite{drwa10a}. It remains invariant under complex permutation of atoms (see Example~\ref{ex:allintsym}).
\end{example}

\section{Group Theoretic Background}
Intuitively, a symmetry of a discrete object is a transformation of its components that leaves the object unchanged. By a symmetry of an answer set program we mean a permutation of its atoms that does not change the logic program, in particular, maps rules to rules. In principle, such a permutation can affect arbitrarily many atoms at once, for instance, as in the case of a complete cyclic shift.

Symmetries are studied in terms of groups. A \emph{group} is an abstract algebraic structure~$(G,\ast)$, where $G$ is a set closed under a binary associative operation~$\ast$ such that there is a \emph{unit} element and every element has a unique \emph{inverse}. Often, we abuse notation and refer to the group~$G$, rather than to the structure $(G,\ast)$. A subset $H$ of $G$ is referred to as a \emph{subgroup} of $G$ if $H$ is closed under the binary operation of $G$. A set of group elements $H \subset G$ such that any other group element in $G$ can be expressed in terms of their product is called a \emph{generating set} and the elements of $H$ are called \emph{generators} of $G$. A generator is \emph{redundant} if it can be expressed in terms of other generators. An \emph{irredundant} generating set does not contain redundant generators, and provides an extremely compact representation of the group. In fact, representing groups by sets of generators always ensures exponential compression.
\begin{theorem}[Exponential Compression \cite{sa09a}]
Any irredundant generating set for a finite group $G$, such that $|G| > 1$, contains at most $\log_2|G|$ elements.
\end{theorem}
A \emph{permutation} of a set~$A$ is a bijection $\pi : A \to A$. Indeed, the set of permutations form a group under composition, denoted as $S(A)$. It is easy to see that the composition of two permutations is a permutation, that the composition of permutations is associative, that the composition with the \emph{identity} never changes a permutation, and that every permutation has a unique inverse.
%\begin{theorem}[Permutation Group]
%Given a non-empty set $A$. The set of all permutations of $A$, denoted as $S(A)$, is a group under composition.
%\end{theorem}

The image of $a \in A$ under a permutation $\pi$ is denoted as $a^\pi$, and for $S \subseteq A$ define $S^\pi = \{a^\pi \mid a \in S\}$. The \emph{orbits} of $S$ under $\pi$ are the set of elements of $A$ to which $S$ can be mapped by (repeatedly) applying $\pi$. Orbits under a permutation define an equivalence relation on $A$. Analogously, for vectors $v = (v_1, v_2, \dots, v_k) \in A^k$ define $v^\pi = (v_1^\pi, v_2^\pi, \dots, v_k^\pi)$, and sets of sets $S = \{S_1, S_2, \dots, S_k\}$ such that $S_i \subseteq A$ for $1 \leq i \leq k$ define $S^\pi = \{S_1^\pi, S_2^\pi, \dots, S_k^\pi\}$.

For a logic program $P$, we define the \emph{symmetric group} of $P$, $S(\atom{P})$, to be the group of all permutations of the atoms that occur in $P$.
We will make use of the \emph{cycle notation} where a permutation is a product of disjoint cycles. A cycle~$(1\ 2\ 3\ \dots\ n)$ means that the permutation maps $1$ to $2$, $2$ to $3$, and so on, finally $n$ back to $1$. An element that does not appear in any cycle is understood as being mapped to itself. Table \ref{fig:graphs} provides some examples. Finally, we define the \emph{support} \cite{mc81a} of a permutation as those elements that are not mapped to themselves.

As to Symmetry-breaking Answer Set Solving, given a logic program $P$, we are interested in the subgroup of the symmetric group of $P$ which elements leave $P$ unchanged. Obviously, a symmetry of a logic program preserves answer sets.
\begin{definition}[Symmetry of a Logic Program]
A symmetry of a logic program $P$ is a permutation of its atoms that does not change $P$.
\end{definition}

\begin{example} \label{ex:allintsym}
There are four symmetries in the All-interval Series problem: (1) the identity, (2) reversing the series (variable symmetry), (3) reflecting the series by substracting each element from $n-1$ (value symmetry), and (4) doing both. It is easy to see that (2) and (3) form a group of generators. Indeed, we can find both symmetries in our encoding (see Example~\ref{ex:allint}) given in cycle notation below.
\[
\begin{array}{ll}
\pi_2 =& (v_{1,0}\ v_{n,0})\ (v_{1,1}\ v_{n,1})\ \dots\ (v_{1,n-1}\ v_{n,n-1})\\
       & \dots \\
       & (v_{\lfloor n/2 \rfloor,0}\ v_{\lceil n/2 \rceil,0})\ (v_{\lfloor n/2 \rfloor,1}\ v_{\lceil n/2 \rceil,1})\ \dots\ (v_{\lfloor n/2 \rfloor,n-1}\ v_{\lceil n/2 \rceil,n-1}) \\
       & (d_{1,1}\ d_{n-1,1})\ (d_{1,2}\ d_{n-1,2})\ \dots\ (d_{1,n-1}\ d_{n-1,n-1}) \\
       & \dots \\
       & (d_{\lfloor (n-1)/2 \rfloor,1}\ d_{\lceil (n-1)/2 \rceil,1})\ (d_{\lfloor (n-1)/2 \rfloor,2}\ d_{\lceil (n-1)/2 \rceil,2})\ \\
       & \dots\ (d_{\lfloor (n-1)/2 \rfloor,n-1}\ d_{\lceil (n-1)/2 \rceil,n-1}) \\ \\
\pi_3 =& (v_{1,0}\ v_{1,n-1})\ (v_{1,1}\ v_{1,n-2})\ \dots\ (v_{n,\lfloor (n-1)/2 \rfloor}\ v_{n,\lceil (n-1)/2 \rceil})\\
       & \dots \\
       & (v_{n,0}\ v_{n,n-1})\ (v_{n,1}\ v_{n,n-2})\ \dots\ (v_{n,\lfloor (n-1)/2 \rfloor}\ v_{n,\lceil (n-1)/2 \rceil})
\end{array}
\]
Intuitively, the circles in the first three lines of $\pi_2$ simply swap the first and the last variable, the second and the last but one variable, etc., value by value to reverse the series, where the remaining circles adjust the auxiliary variables, i.e., swap the differences value by value, respectively. The circles in $\pi_3$ swap the values $0$ and $n-1$, $1$ and $n-2$, etc., for each variable to reflect the series. Obviously, the permutations $\pi_2$ and $\pi_3$ represent (2) and (3), respectively, and do not change the logic program.
\end{example}

\section{ASP Symmetries via Graph Automorphism \label{sec:sd}}
Our approach for detecting symmetries of a logic program is through reduction to, and solution of, an associated Graph Automorphism problem (GAP).

Given a graph $G = (V,E)$, where $V$ is a set of vertices and $E \subseteq V \times V$ is a set of edges. An \emph{automorphism} (symmetry) of $G$ is a permutation of its vertices that maps edges to edges, and non-edges to non-edges. Edge orientation must be preserved in case $G$ is a directed graph. A further extension considers vertex colourings, where symmetries must map each vertex into a vertex with the same colour. More formally, given a partition of the vertices $\pi(V) = \{V_1, V_2, \dots, V_k\}$, the \emph{automorphism group} of $G$ \cite{mc81a} is $Aut(G,\pi) = \{ \gamma \in S(V) \mid (V^\gamma, E^\gamma) = (V,E) , \pi^\gamma = \pi\}$. We will think of the partition $\pi$ as a \emph{colouring} of the vertices.
The \emph{(Coloured) Graph Automorphism problem} (GAP) is to find all symmetries of a given graph, for instance, in terms of generators. It is not known to have any polynomial time solution, and is conjectured to be strictly between the complexity classes P and NP \cite{ba95a}, thus potentially easier than computing answer sets. A practical algorithm for graph automorphism has been implemented in \systemname{nauty}~\cite{mc81a} and significantly improved in the systems \systemname{saucy}~\cite{dalisama04a,dasakama08a}.

A quite natural GAP encoding for detecting symmetries of logic programs is based on their body-atom graph. The \emph{body-atom graph} $G_P = (V,E_0 \cup E_1,E_2)$ of a logic program $P$ is a directed graph with vertices $V = \body{P} \cup \atom{P}$, and labelled edges $E_0 = \{(\beta, a) \mid a \in \atom{P} , \beta \in \body{a}\}$, $E_1 = \{(a,\beta) \mid \beta \in \body{P} , a \in \beta^+\}$, and $E_2 = \{(a,\beta) \mid \beta \in \body{P} , a \in \beta^-\}$.
The body-atom graph has been shown to be a suitable representation of a logic program \cite{linke03b}. However, we modify the body-atom graph by introducing additional vertices for negated atoms to circumvent labelled edges.
\begin{figure}
\begin{center}
\begin{tabular}{c@{\hspace{5em}}c}
\begin{tikzpicture}
	[circle, inner sep=0pt, minimum size=5.0mm, >=stealth]
	\node (b)	at ( 0.0, 0.0) [draw] {$\beta$};
	\node (a1)	at (-1.5, 0.5) [draw] {$\nodelabel{a}{1}$};
	\node (al)	at (-1.5,-0.5) [draw] {$\nodelabel{a}{l}$};
	\node (b1)	at ( 1.5, 1.5) [draw] {$\nodelabel{b}{1}$};
	\node (bm)	at ( 1.5, 0.5) [draw] {$\nodelabel{b}{m}$};
	\node (nc1)	at ( 1.5,-0.5) [draw] {};
	\node (nc1)	at ( 1.5,-0.5) [] {$\dneglabel{c}{1}$};
	\node (ncn)	at ( 1.5,-1.5) [draw] {};
	\node (ncn)	at ( 1.5,-1.5) [] {$\dneglabel{c}{n}$};
	\draw [-, dotted, thick] (-1.5, 0.125) -- (-1.5,-0.125);
	\draw [-, dotted, thick] ( 1.5, 1.125) -- ( 1.5, 0.875);
	\draw [-, dotted, thick] ( 1.5,-1.125) -- ( 1.5,-0.875);
	\draw [->] (b) -- (a1);
	\draw [->] (b) -- (al);
	\draw [->] (b1) -- (b);
	\draw [->] (bm) -- (b);
	\draw [->, snake=snake, segment amplitude=0.5mm, segment length=2.5mm, line after snake=0.5mm, line before snake=0.5mm] (nc1) -- (b);
	\draw [->, snake=snake, segment amplitude=0.5mm, segment length=2.5mm, line after snake=0.5mm, line before snake=0.5mm] (ncn) -- (b);
\end{tikzpicture} &
\begin{tikzpicture}
	[circle, inner sep=0pt, minimum size=5.0mm, >=stealth]
	\node at ( 1.5,-0.5) [fill, color=lightgray] {};
	\node at ( 1.5,-1.5) [fill, color=lightgray] {};
	\node at (-3.0, 0.5) [fill, color=lightgray] {};
	\node at (-3.0,-0.5) [fill, color=lightgray] {};
	\node at ( 3.0, 1.5) [fill, color=lightgray] {};
	\node at ( 3.0, 0.5) [fill, color=lightgray] {};
	\node (b)	at ( 0.0, 0.0) [draw, rectangle] {$\beta$};
	\node (a1)	at (-1.5, 0.5) [draw] {$\nodelabel{a}{1}$};
	\node (al)	at (-1.5,-0.5) [draw] {$\nodelabel{a}{l}$};
	\node (na1)	at (-3.0, 0.5) [draw] {};
	\node 		at (-3.0, 0.5) [] {$\dneglabel{a}{1}$};
	\node (nal)	at (-3.0,-0.5) [draw] {};
	\node 		at (-3.0,-0.5) [] {$\dneglabel{a}{l}$};
	\node (b1)	at ( 1.5, 1.5) [draw] {$\nodelabel{b}{1}$};
	\node (nb1)	at ( 3.0, 1.5) [draw] {};
	\node 		at ( 3.0, 1.5) [] {$\dneglabel{b}{1}$};
	\node (bm)	at ( 1.5, 0.5) [draw] {$\nodelabel{b}{m}$};
	\node (nbm)	at ( 3.0, 0.5) [draw] {};
	\node 		at ( 3.0, 0.5) [] {$\dneglabel{b}{m}$};
	\node (nc1)	at ( 1.5,-0.5) [draw] {};
	\node 		at ( 1.5,-0.5) [] {$\dneglabel{c}{1}$};
	\node (ncn)	at ( 1.5,-1.5) [draw] {};
	\node 		at ( 1.5,-1.5) [] {$\dneglabel{c}{n}$};
	\node (c1)	at ( 3.0,-0.5) [draw] {$\nodelabel{c}{1}$};
	\node (cn)	at ( 3.0,-1.5) [draw] {$\nodelabel{c}{n}$};
	\draw [-, dotted, thick] (-1.5, 0.125) -- (-1.5,-0.125);
	\draw [-, dotted, thick] ( 1.5, 1.125) -- ( 1.5, 0.875);
	\draw [-, dotted, thick] ( 1.5,-1.125) -- ( 1.5,-0.875);
	\draw [-, dotted, thick] (-3.0, 0.125) -- (-3.0,-0.125);
	\draw [-, dotted, thick] ( 3.0, 1.125) -- ( 3.0, 0.875);
	\draw [-, dotted, thick] ( 3.0,-1.125) -- ( 3.0,-0.875);
	\draw [->] (a1) -- (na1);
	\draw [->] (al) -- (nal);
	\draw [->] (b) -- (a1);
	\draw [->] (b) -- (al);
	\draw [->] (b1) -- (b);
	\draw [->] (bm) -- (b);
	\draw [->] (b1) -- (nb1);
	\draw [->] (bm) -- (nbm);
	\draw [->] (nc1) -- (b);
	\draw [->] (ncn) -- (b);
	\draw [->] (c1) -- (nc1);
	\draw [->] (cn) -- (ncn);
\end{tikzpicture}
\end{tabular}
\end{center}
\caption{The left picture shows a rule $r$ of the form (\ref{form:rule}) as a body-atom-graph, where $\beta$ is the body vertex. Straight lines represent edges in $E_0 \cup E_1$, curly lines represent edges in $E_2$. On the right is the general structure of a 3-coloured graph construction of $r$. Vertices of color 1, 2, and 3 are represented by empty circles, filled circles, and empty squares, respectively.}
\end{figure}
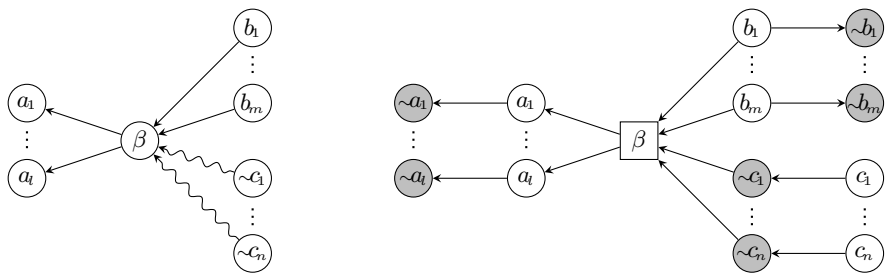

\begin{figure}
\begin{center}
\begin{tabular}{c@{\hspace{5em}}c}
\begin{tikzpicture}
	[circle, inner sep=0pt, minimum size=5.0mm, >=stealth]
	\node at (0.0,0.0) [fill, color=lightgray] {};
	\node at (3.0,1.5) [fill, color=lightgray] {};
	\node (lna) at (0.0,0.0) [draw] {$\dneg a$};
	\node (la)  at (0.0,1.5) [draw] {$a$};
	\node (b2)  at (1.5,0.0) [rectangle, draw] {2};
	\node (b1)  at (1.5,1.5) [rectangle, draw] {1};
	\node (lb)  at (3.0,0.0) [draw] {$b$};
	\node (lnb) at (3.0,1.5) [draw] {$\dneg b$};
	\draw [->] (lna) -- (b2);
	\draw [->] (b2) -- (lb);
	\draw [->] (lnb) -- (b1);
	\draw [->] (b1) -- (la);
	\draw [->] (la) -- (lna);
	\draw [->] (lb) -- (lnb);
\end{tikzpicture} & 
\begin{tikzpicture}
	[circle, inner sep=0pt, minimum size=5.0mm, >=stealth]
	\node at (0.0,0.0) [fill, color=lightgray] {};
	\node at (4.5,1.5) [fill, color=lightgray] {};
	\node (lna) at (0.0,0.0) [draw] {$\dneg a$};
	\node (la)  at (1.5,0.0) [draw] {$a$};
	\node (b1)  at (1.5,1.5) [rectangle, draw] {1};
	\node (b2)  at (3.0,0.0) [rectangle, draw] {2};
	\node (lb)  at (3.0,1.5) [draw] {$b$};
	\node (lnb) at (4.5,1.5) [draw] {$\dneg b$};
	\draw [->] (la) -- (b2);
	\draw [->] (b1) -- (lb);
	\draw [->] (lb) -- (b2);
	\draw [->] (b1) -- (la);
	\draw [->] (la) -- (lna);
	\draw [->] (lb) -- (lnb);
\end{tikzpicture} \\
Original 3-coloured graph of $P_1$ & Original 3-coloured graph of $P_2$ \\ \noalign{\vspace {.5cm}}
\begin{tikzpicture}
	[circle, inner sep=0pt, minimum size=5.0mm, >=stealth]
	\node at (0.0,0.0) [fill, color=lightgray] {};
	\node at (3.0,1.5) [fill, color=lightgray] {};
	\node (lna) at (0,0) [draw] {$\dneg b$};
	\node (la)  at (0,1.5) [draw] {$b$};
	\node (b2)  at (1.5,0) [rectangle, draw] {1};
	\node (b1)  at (1.5,1.5) [rectangle, draw] {2};
	\node (lb)  at (3,0) [draw] {$a$};
	\node (lnb) at (3,1.5) [draw] {$\dneg a$};
	\draw [->] (lna) -- (b2);
	\draw [->] (b2) -- (lb);
	\draw [->] (lnb) -- (b1);
	\draw [->] (b1) -- (la);
	\draw [->] (la) -- (lna);
	\draw [->] (lb) -- (lnb);
\end{tikzpicture} &
 \begin{tikzpicture}
	[circle, inner sep=0pt, minimum size=5.0mm, >=stealth]
	\node at (0.0,0.0) [fill, color=lightgray] {};
	\node at (4.5,1.5) [fill, color=lightgray] {};
	\node (lna) at (0.0,0.0) [draw] {$\dneg b$};
	\node (la)  at (1.5,0.0) [draw] {$b$};
	\node (b1)  at (1.5,1.5) [rectangle, draw] {1};
	\node (b2)  at (3.0,0.0) [rectangle, draw] {2};
	\node (lb)  at (3.0,1.5) [draw] {$a$};
	\node (lnb) at (4.5,1.5) [draw] {$\dneg a$};
	\draw [->] (la) -- (b2);
	\draw [->] (b1) -- (lb);
	\draw [->] (lb) -- (b2);
	\draw [->] (b1) -- (la);
	\draw [->] (la) -- (lna);
	\draw [->] (lb) -- (lnb);
\end{tikzpicture} \\
$\pi_1 = (a\ b)\ (\dneg a\!\dneg b)\ (1\ 2)$ & $\pi_2 = (a\ b)\ (\dneg a\!\dneg b)$\\
\end{tabular}
\end{center}
\caption{3-coloured graph constructions and resulting symmetries for the example logic programs $P_1$ and $P_2$. \label{fig:graphs}}
\end{figure}
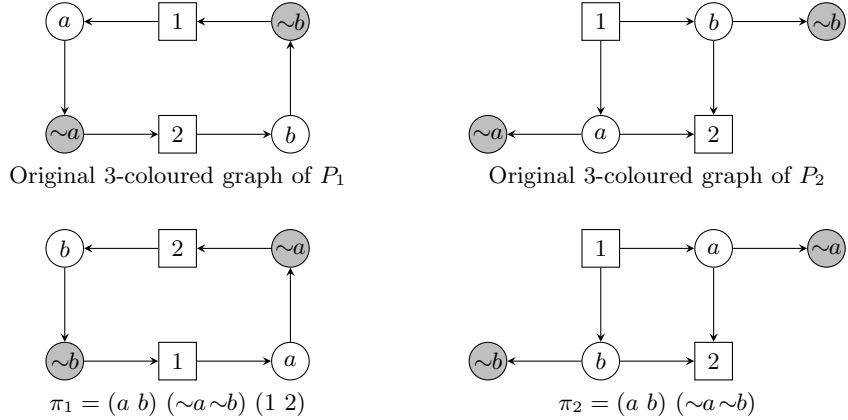

In our GAP encoding every atom in $\atom{P}$ is represented by two vertices of colour $1$ and $2$ that correspond to the positive and negative literals, respectively. Every rule is represented by a (body) vertex of colour $3$, a set of directed edges that connect the vertices of the literals that appear in the rule's body to its body vertex, and a set of directed edges that connect the body vertex to the vertices of the atoms (positive literals) that appear in the head of the rule. To ensure consistency, that is, $a$ maps to $b$ if and only if $\dneg a$ maps to $\dneg b$ for any atoms $a$ and $b$, vertices of opposite literals are mated by a directed edge from the positive literal to the negative literal. The choice of three vertex colours insures that body vertices can only be mapped to body vertices, and positive (negative) literal vertices can only be mapped to positive (negative) literal nodes. To conclude, given a logic program $P$ consisting of $m$ rules and $l$ literals over $n$ atoms, the GAP encoding for detecting symmetries of $P$ is constructed by $m+2n$ vertices and $l+n$ edges. Examples are given in Fig. \ref{fig:graphs}.

Since graph automorphism algorithms are sensitive to the number of vertices of an input graph, our construction can be optimised to reduce the number of graph vertices while preserving its automorphism group. A first simplification is achieved by modelling rules with an empty body and a single head atom, so-called \emph{facts}, by a (forth) colour for the vertex corresponding to the head atom instead of using (empty) body vertices. Furthermore, rules with a single head atom and a 1-literal body are modelled using a directed edge from the vertex corresponding to the literal of the body to the vertex corresponding to the head atom. Observe that this optimisation may connect a literal vertex to a positive literal vertex, where consistency edges connect positive literal vertices to their negative mates. Therefore, unintended mappings between 1-literal body edges and consistency edges are impossible. For the special case of a 1-literal body and an empty head, we connect the literal vertex to the special node '$\bot$'.

\section{Symmetry-breaking Constraints \label{sec:sb}}
Recall that symmetries of a logic program $P$ induce equivalence classes in the solution space (orbits). Given an answer set of $P$, all sets to which it can be mapped by symmetries, must be answer sets of $P$. Similarly, symmetries always map non-answer sets to non-answer sets. Therefore, it is sufficient to reason about one representative from every equivalence class. Symmetry-breaking amounts to selecting some representatives from every equivalence class and constructing rules, composed into a symmetry-breaking constraint, that is only satisfied on those representatives. A \emph{full} SBC selects exactly one representative from each orbit, otherwise we call an SBC \emph{partial}. The most common approach is to order all elements from the solution space lexicographically, and to select the lexicographically smallest element, the \emph{lex-leader}, from each orbit as its representative (cf. \cite{crgiluro96a,almasa03a,alramasa03a,sa09a}). A \emph{lex-leader symmetry-breaking constraint} is an SBC that is satisfied only on the lex-leaders of orbits.

We will assume a total ordering on the atoms $a_1, a_2, \dots, a_n$ of a logic program $P$ and consider the induced lexicographic ordering on the truth assignments, i.e. their interpretation as unsigned integers. The most common approach for accomplishing the construction of a lex-leader SBC is by encoding a \emph{permutation constraint} (PC) for every permutation $\pi$, where
\[
PC(\pi) = \bigwedge_{1 \leq i \leq n} \left\lbrack \bigwedge_{1 \leq j \leq i-1} (a_j = {a_j}^\pi) \right\rbrack \rightarrow(a_i \leq {a_i}^\pi).
\]
A careful analysis reveals some possibilities to reduce the size of permutation constraints (cf. \cite{sa09a}). The first corresponds to atoms that are mapped to themselves by the permutation, i.e., ${a_i}^\pi = a_i$. This makes the consequent of the implication unconditionally true. For sparse symmetries, one can significantly reduce the size of the permutation constraint with a restriction of the PC construction to only those atoms that are in the support of $\pi$. A second possibility corresponds to the lexicographically biggest atom in each cycle of $\pi$. Assume a cycle $(a_s \dots a_e)$ on the atoms of some index set $J$. Using equality propagation on the portion of the permutation constraint where $i=e$, we get $(a_s = a_e) \rightarrow (a_e \leq a_s)$ which is tautologous. Hence, we can further restrict the index set in the PC by excluding the lexicographically biggest atom in each cycle.

Through \emph{chaining} which includes additional atoms, we achieve a PC representation that is linear in the number of atoms (cf. \cite{almasa03a}):
\[
\begin{array}{l@{\ \equiv\ }l}
c_{\pi,1} & (a_1 \leq a_1^\pi) \land c_{\pi,2}\\
c_{\pi,i} & (a_{i-1} \geq a_{i-1}^\pi) \rightarrow (a_i \leq a_i^\pi) \land c_{\pi,i+1}\qquad i = 2,\dots,n\\
c_{\pi,n+1} & 1
\end{array}
\]
Finally, we encode above permutation constraint in ASP that is satisfied for the lex-leader in the orbit induced by $\pi$ as follows
\[
\begin{array}{r@{\ \leftarrow\ }l}
& a_1, \dneg {a_1}^\pi \\
& c_{\pi,2} \\
c_{\pi,i} & a_{i-1}, a_i, \dneg {a_i}^\pi \\
c_{\pi,i} & a_i, \dneg {a_{i-1}}^\pi, \dneg {a_i}^\pi \\
c_{\pi,i} & a_{i-1}, c_{\pi,i+1} \\
c_{\pi,i} & c_{\pi,i+1}, \dneg {a_{i-1}}^\pi \\
c_{\pi,n+1} &
\end{array}
\]
where $i=2,\dots,n$. The lex-leader symmetry-breaking constraint that breaks every symmetry in a logic program can now be constructed by conjoining all of its permutation constraints. 
\begin{example}
We illustrate our PC encoding on the symmetries detected for the previous examples $P_1$ and $P_2$. Since both permutations $\pi_1$ and $\pi_2$ (see Fig. \ref{fig:graphs}) map $a$ to $b$ and vice versa, they share the same lex-leader SBC which is as simple as follows (assuming $a$ is lexicographically greater than $b$):
\[
\leftarrow b, \dneg a
\]
Observe that the ordering on the atoms of a logic program $P$ induces a preference relation on the answer sets of $P$ under symmetry-breaking. Here, the ordering selects $\{a\}$ as the representative of the set of all answer sets symmetric to $\{a\}$, hence, eliminating the answer set $\{b\}$. 
\end{example}

\section{Partial Symmetry-breaking \label{sec:psb}}
Breaking all symmetries may not speed up search because there are often exponentially many of them. A better trade-off may be provided by breaking \emph{enough} symmetries \cite{crgiluro96a}. Irredundant generators are good candidates because they can not be expressed in terms of each other, and implicitly represent all symmetries.
Hence, breaking all symmetries in a generating set can eliminate all problem symmetries. However, this does not hold in general. In fact, it has been shown that even when breaking all symmetries is polynomial, there exists cases where for which SBCs based on any irredundant generating set fail to break all symmetry \cite{kanawa09a}.

We can further reduce the size of symmetry-breaking constraints by restricting the construction of permutation constraints up to the $k$-th atom in each permutation \cite{almasa03a}.
\begin{example}
Consider the All-interval Series problem encoded as in Example~\ref{ex:allint} and the generators $\pi_2$ and $\pi_3$ from Example~\ref{ex:allintsym}.
The symmetry-breaking constraint, where both permutation constraints are restricted to the second atom, is given through the following, where $c_0, \dots, c_3$ are new atoms.
\[
\begin{array}{r@{\ \leftarrow\ }l@{\qquad}r@{\ \leftarrow\ }l}
& v_{1,0}, \dneg v_{1,n-1}                      & & v_{1,0}, \dneg v_{n,0} \\
& c_0                                           & & c_2\\
c_0 & v_{1,0}, v_{1,1}, \dneg v_{1,n-2}         & c_2 & v_{1,0}, v_{1,1}, \dneg v_{n,1} \\
c_0 & v_{1,1}, \dneg v_{1,n-1}, \dneg v_{1,n-2} & c_2 & v_{1,1}, \dneg v_{n,0}, \dneg v_{n,1}\\
c_0 & v_{1,0}, c_1                              & c_2 & v_{1,0}, c_3 \\
c_0 & c_1, \dneg v_{1,n-1}                      & c_2 & c_3, \dneg v_{n,0}\\
c_1 &                                           & c_3 & \\
\end{array}
\]
\end{example}

\section{Experiments \label{sec:exp}}
Our approach to Symmetry-breaking Answer Set Solving has been implemented within the preprocessor \systemname{sbass}\footnote{\texttt{http://potassco.sourceforge.net/} provides \systemname{clasp}, \systemname{gringo}, and \systemname{sbass}}. It accepts a logic program in \systemname{smodels}\footnote{\texttt{http://www.tcs.hut.fi/Software/smodels/} provides \systemname{lparse} and \systemname{smodels}} format~\cite{lparseManual} produced by a grounder, e.g. \systemname{lparse} and \systemname{gringo}, and incorporates the graph automorphism tool \systemname{saucy}\footnote{\texttt{http://vlsicad.eecs.umich.edu/BK/SAUCY/}} (2.1) for detecting irredundant generators of the group of permutational symmetries. In return, \systemname{sbass} outputs the given program together with symmetry-breaking constraints, again in \systemname{smodels} format, which can be applied to any suitable answer set solver, e.g. \systemname{smodels} and \systemname{clasp}. Note that \emph{sbass} provides several options, for instance, to print detected generators in cycle notation or statistics.

To evaluate our approach, we conducted experiments on ASP encodings of several difficult combinatorial search problems. Experiments consider the answer set solver \systemname{clasp} (1.3.2) on instances with symmetry-breaking in terms of generators, i.e., instances preprocessed by \systemname{sbass}, and without symmetry-breaking. To explore the impact of partial SBC, we also tried restrictions on the construction of permutation constraints up to the $k$-th support in a permutation, denoted as \systemname{clasp}$_k^\pi$.
All tests were run on a 2.00~GHz PC under Linux, where each run was limited to 600 s time and 1 GB RAM, preprocessing excluded. However, we also report the runtime for \systemname{sbass} and give the number of generators. The latter gives an impression about the size of the search space implicitly pruned through symmetry-breaking.
In the experiments below we generally compare the runtime for testing the existence of an answer set to a given problem.

\subsection{Pigeon Hole Problems}
The \emph{Pigeon Hole problem} is to show that it is impossible to put $n$ pigeons into $n-1$ holes if each pigeon must be put into a distinct hole. This problem is provably exponentially hard for any resolution based method, but is tractable using symmetries (all the pigeons are interchangeable and all the holes are interchangeable). We encoded the Pigeon Hole problem based on the support encoding for the all-different constraint \cite{drwa10a}, as follows, where $p_{ij}$ is taken to mean that pigeon $i$ is assigned hole $j$:
\[
\begin{array}{r@{\ \leftarrow\ }l@{\qquad}l}
p_{i,1} ; p_{i,2} ; \dots ; p_{i,n-1} &  & i \in 1..n\\
& p_{i,j}, p_{k,j} & i \neq k \\
\end{array}
\]
The runtimes for various sizes of $n$ are shown in Table \ref{tab:php}. Although symmetry-breaking has a positive impact, the runtime even with \systemname{clasp}$_\infty^\pi$ is still exponentially growing with the number of pigeons. Here, symmetry-breaking on the generating set returned by \systemname{saucy} does not break all problem symmetries. We enforced \systemname{saucy} to compute a different set of generators, denoted as \systemname{clasp}$_\infty^\theta$, and got a polynomial runtime.
On such problems, full SBCs are essential.
\begin{table}
\caption{Runtime results in seconds for Pigeon Hole problems. \label{tab:php}}
\centering
\begin{tabular}{ccccccccc}
\hline\noalign{\smallskip}
\#$n$&\#gen.& \systemname{sbass} & \systemname{clasp}$_1^\pi$ & \systemname{clasp}$_{5}^\pi$ & \systemname{clasp}$_\infty^\pi$ & \systemname{clasp}$_\infty^\theta$ & \systemname{clasp}\\  
\noalign{\smallskip}
\hline
\noalign{\smallskip}
11 & 18 & 0.05 &  0.38 &  0.15 &  0.06 & 0.02 &  0.62 \\
12 & 20 & 0.08 &  4.09 &  0.07 &  0.22 & 0.03 &  5.99 \\
13 & 22 & 0.11 & 30.57 &  0.43 &  0.32 & 0.03 & 53.39 \\
14 & 24 & 0.16 &272.72 &  4.95 &  1.73 & 0.04 &448.98 \\
15 & 26 & 0.23 & ---   & 62.61 &  3.02 & 0.04 & ---   \\
16 & 28 & 0.32 & ---   & ---   & 23.01 & 0.07 & ---   \\
17 & 30 & 0.44 & ---   & ---   &130.87 & 0.10 & ---   \\
\hline
\end{tabular}
\end{table}

\subsection{Ramsey's Theorem}
Ramsey's Theorem states that for any pair of positive integers $(k,m)$ there exists a least positive integer $n$ such that, no matter how we color the edges of the clique with $n$ vertices, $K_n$, using two colours, say blue and red, there is a sub-clique with $k$ nodes of colour blue or a sub-clique with $m$ nodes of colour red. We used the encoding from \cite{lepffaeigopesc02a}, denoted as $R(k,m,n)$, to determine whether $n$ is not an integer for which the theorem holds.

In formerly hard cases, namely $R(3,5, 14)$ and $R(4,5, 24)$, symmetry-breaking lead to significant pruning of the search space and yield solutions in a considerably short amount of time. The results presented in Table \ref{tab:ramsey} suggest full SBCs for unsatisfiable instances, but small, partial SBCs for satisfiable instances.

\begin{table}
\caption{Average time for completed runs in seconds and the number of timeouts on Ramsey's Theorem instances, each shuffled 5 times. The $^\ast$asterisk denotes instances that have no answer sets.\label{tab:ramsey}}
\centering
\begin{tabular}{lcccccccccc}
\hline\noalign{\smallskip}
 & & \systemname{sbass} & \multicolumn{2}{c}{\systemname{clasp}$_1^\pi$} & \multicolumn{2}{c}{\systemname{clasp}$_5^\pi$} & \multicolumn{2}{c}{\systemname{clasp}$_\infty^\pi$} & \multicolumn{2}{c}{\systemname{clasp}}\\ 
   & \#gen. & time & time & \#t.out & time & \#t.out & time & \#t.out & time & \#t.out\\
\noalign{\smallskip}
\hline
\noalign{\smallskip}
$R(3,5, 13)$ & 11 & 0.06 & \textbf{0.01} &  &  \textbf{0.01} &  &  0.03 &  &  \textbf{0.01} &  \\
$R(3,5, 14)^\ast$ & 12 & 0.10 &  3.58 &  &  1.23 &  &  \textbf{0.49} &  &354.25 &  \\
$R(3,6, 17)$ & 15 & 1.18 &  0.12 &  &  0.12 &  &  0.14 &  &  \textbf{0.11} &  \\
$R(3,6, 18)^\ast$ & 16 & 1.87 & ---   & 5&  ---  & 5& ---   & 5& ---   & 5\\
$R(4,4, 17)$ & 15 & 0.26 &  0.73 &  &  0.12 &  &  0.50 &  &  \textbf{0.07} &  \\
$R(4,4, 18)^\ast$ & 16 & 0.37 & ---   & 5&  ---  & 5& ---   & 5& ---   & 5\\
$R(4,5, 23)$ & 21 & 5.43 &  4.23 &  &  2.29 &  &  2.05 &  &  \textbf{1.32} &  \\
$R(4,5, 24)$ & 22 & 7.15 & \textbf{77.64} &  &208.66 & 1&180.96 & 3& ---   & 5\\
$R(4,5, 25)^\ast$ & 23 & 9.54 & ---   & 5& ---   & 5& ---   & 5& ---   & 5\\
\hline
\end{tabular}
\end{table}

\subsection{Graceful Graphs}
A labelling $f$ of the vertices of a graph $(V,E)$ is \emph{graceful} if $f$ assigns a unique label~$f(v)$ from $\{0,1,\dots,|E|\}$ to each vertex $v \in V$ such that, when each edge $(v,w) \in E$ is assigned the label $|f(v)-f(w)|$, the resulting edge labels are distinct. The problem of determining the existence of a graceful labelling of a graph has been modelled as a CSP in \cite{pesm03a}, and is an interesting application for Symmetry-breaking Answer Set Solving because the symmetries are different for each instance and can not be modelled a-priori in general. Our experiments consider graphs $DW_n$ and $K_nP_m$. The \emph{double wheel} graph $DW_n$ is composed of two copies of a cycle with $n$ vertices, each connected to a central hub. The two wheels $W_n$, each have rotation and reflection symmetries. The labels of the two cycles can also be interchanged. The graph $K_nP_m$ is the cross-product of the clique $K_n$ and the path $P_m$. It consists of $n$ copies of $K_n$, with corresponding vertices in the $m$ cliques also forming the vertices of a path $P_m$. Symmetries of the graph are simultaneous rotations of the cliques and inter-clique permutations.

As can be seen in Table \ref{tab:graceful}, we achieve speed-up on the unsatisfiable instance $DW_3$. For the other instances, all of which are satisfiable, the branching heuristic used in our approach sometimes appears to be misled by the extra variables introduced in \systemname{clasp}$_k^\pi$. That explains some of the variability in the runtimes. However, the difficult instances show symmetry-breaking to be outperforming.
\begin{table}
\caption{Average time for completed runs in seconds and the number of timeouts on Graceful Graph instances, each shuffled 5 times. The $^\ast$asterisk denotes instances that have no answer sets.\label{tab:graceful}}
\centering
\begin{tabular}{lcccccccccc}
\hline\noalign{\smallskip}
 & & \systemname{sbass} & \multicolumn{2}{c}{\systemname{clasp}$_1^\pi$} & \multicolumn{2}{c}{\systemname{clasp}$_5^\pi$} & \multicolumn{2}{c}{\systemname{clasp}$_\infty^\pi$} & \multicolumn{2}{c}{\systemname{clasp}}\\ 
   & \#gen. & time & time & \#t.out & time & \#t.out & time & \#t.out & time & \#t.out\\
\noalign{\smallskip}
\hline
\noalign{\smallskip}
${DW_3}^\ast$   & 5 & 0.02 & 4.24 & & 1.45 & & \textbf{1.32} & & 5.40 &\\
%$DW_4$   & 5 & 0.05 & 0.10 & & \textbf{0.09} & & 0.26 & & 0.14 &\\
$DW_6$   & 5 & 0.17 & \textbf{0.46} & & 0.56 & & 1.09 & & 0.57 &\\
$DW_8$   & 5 & 0.48 &28.81 & & 5.47 & &17.11 & & \textbf{4.30} &\\
$DW_{10}$& 5 & 1.21 &191.86& &66.18 & &\textbf{61.59} & &27.04 &2\\
$DW_{12}$& 5 & 3.34 & \textbf{145.89}& &202.18&1&111.96 &1&112.38 &4\\
$K_3P_3$ & 3 & 0.04 & 0.08 & & 0.08 & & \textbf{0.07} & & 0.08 &\\
$K_4P_2$ & 4 & 0.07 & 0.20 & & \textbf{0.10} & & 0.54 & & 0.19 &\\
$K_4P_3$ & 4 & 0.29 &24.68 & &29.06 & &198.57& &\textbf{24.01}& \\
$K_5P_2$ & 5 & 0.37 &274.85&3&334.55&3&\textbf{312.56}&\textbf{1}&226.03&3\\
\hline
\end{tabular}
\end{table}

\subsection{Answer Set Enumeration \label{sec:enum}}
Finally, we want to test the impact of symmetry-breaking on the number of answer sets. We modelled the All-interval Series problem (AllInt) as described in Example~\ref{ex:allint}, using a direct representation for $n$ integer variables and auxiliary variables to represent the differences between adjacent numbers (cf.~\cite{drwa10a}), and required both sets of variables to be all-different.

As expected we observe that symmetry-breaking significantly reduces the number of solutions, and therefore, reduces the time necessary for post-processing solutions (see Table~\ref{tab:models}). Clearly, \systemname{clasp}$_k^\pi$ for an increasing number~$k$ discards more solutions (eliminating up to 90 per cent of the solution space).

As in all previous benchmarks, it seems safe to assume that the detection of symmetries in logic programs through reduction to graph automorphism is computationally quite feasible using today's GAP tools such as \systemname{saucy}.

We should also note that a given problem can be encoded in many equivalent logic programs, and with each different encoding our techniques may detect a different generating set. For instance, we tried symmetry detection and symmetry-breaking on logic programs that were preprocessed, i.e. simplified. The key idea of preprocessing logic programs is to identify equivalences among its relevant constituents. These equivalences are then used for building a compact representation of the program \cite{gekanesc08a}. Sometimes, we observed significant better results in terms of time and number of answer sets (eliminating up to 95 per cent of the solution space).
\begin{table}
\caption{Results on computing all answer sets of selected instances. Runtime, number of solutions, and the maximum compression achieved using full SBC are shown.\label{tab:models}}
\centering
\begin{tabular}{lccccccccccc}
\hline\noalign{\smallskip}
 & & \systemname{sbass} & \multicolumn{2}{c}{\systemname{clasp}$_1^\pi$} & \multicolumn{2}{c}{\systemname{clasp}$_5^\pi$} & \multicolumn{2}{c}{\systemname{clasp}$_\infty^\pi$} & \multicolumn{2}{c}{\systemname{clasp}} & \\
  & \#gen. & time & time & \#sol. & time & \#sol. & time & \#sol. & time & \#sol. & compr.\\
\noalign{\smallskip}
\hline
\noalign{\smallskip}
$AllInt_{8}$  & 2 & 0.01 &   0.15 & 39   &   0.11 & 15   &   0.17 & 14   &   0.14 & 40    &65\%\\
$AllInt_{9}$  & 2 & 0.01 &   0.78 & 119  &   0.60 & 60   &   0.93 & 40   &   0.77 & 120   &67\%\\
$AllInt_{10}$ & 2 & 0.01 &   4.60 & 295  &   3.43 & 148  &   5.69 & 107  &   4.08 & 296   &64\%\\
$AllInt_{11}$ & 2 & 0.01 &  23.26 & 647  &  22.82 & 372  &  32.70 & 238  &  24.40 & 648   &63\%\\
$AllInt_{12}$ & 2 & 0.01 & 161.90 & 1327 & 147.17 & 862  & 211.27 & 442  & 160.32 & 1328  &67\%\\
$DW_4$        & 5 & 0.07 & 282.36 & 9472 & 168.03 & 5152 &  85.65 & 1150 & 314.15 & 11264 &90\%\\
$K_3P_3$      & 3 & 0.05 & 229.15 & 5704 & 119.99 & 2836 & 126.25 & 1487 & 268.80 & 6816  &76\%\\
$K_4P_2$      & 4 & 0.08 & 119.66 & 1080 &  67.96 & 552  &  27.72 & 146  & 145.13 & 1440  &90\%\\
\hline
\end{tabular}
\end{table}

\section{Conclusions \label{sec:con}}
We have investigated symmetry-breaking in the context of Answer Set Programming. In particular, we proposed a reduction from symmetry detection of disjunctive logic programs to the automorphisms of a coloured digraph. Our techniques were formulated as a completely automated flow that (1) starts with a logic program, (2) detects all of its permutational symmetries, (3) represents all symmetries implicitly and always with exponential compression, (4) adds symmetry-breaking constraints that do not affect the existence of answer sets, and (5) can be applied to any existing ASP system without changing its code, which allows for programmers to select the solvers that best fit their needs.

We have empirically evaluated symmetry-breaking on difficult combinatorial search problems and got promising results. In many cases, SBC lead to significant pruning of the search space and yield solutions to problems which are otherwise intractable.
We also observe a significant compression of the solution space which makes symmetry-breaking attractive whenever all solutions have to be post-processed.

Motivated by this success future work concerns an extension to choice rules and weight constraints \cite{siniso02a}.
However, one should not expect Symmetry-breaking Answer Set Solving to give improvement on all benchmark classes. Many ASP benchmarks\footnote{\texttt{http://asparagus.cs.uni-potsdam.de/}} have large numbers of symmetries, but can be solved so quickly that the symmetry detection and -breaking overhead is not justified.

Furthermore, it is often reasonable to assume that the symmetries for a problem are known. For particular symmetries, there are more efficient breaking methods (cf. \cite{wa06a}). This is also target to future work.

\paragraph*{Acknowledgements} The work of Christian Drescher is partially funded by the Austrian Science Fund (FWF) under grant number P20841 and the Vienna Science and Technology Fund (WWTF) under grant ICT08-020.

\end{document}